# The variational Laplace approach

# to approximate Bayesian inference

*J. Daunizeau[1,2]*

[1] Brain and Spine Institute, Paris, France

[2] ETH, Zurich, France

Address for correspondence:

Jean Daunizeau

Motivation, Brain and Behaviour Group

Brain and Spine Institute (ICM), INSERM UMR S975.

47, bvd de l'Hopital, 75013, Paris, France.

Tel: +33 1 57 27 43 26

Fax: +33 1 57 27 47 94

Mail: jean.daunizeau@gmail.com

Web: https://sites.google.com/site/jeandaunizeauswebsite/

Variational approaches to approximate Bayesian inference provide very efficient means of performing parameter estimation and model selection (Attias, 1999; Beal, 2003). Among these approaches, so-called variational-Laplace or VL schemes rely on Gaussian approximations to posterior densities on model parameters whose impact on the data is nonlinear (Daunizeau et al., 2009; Friston et al., 2007). They owe their popularity to the reasonable computational cost that they entail, especially when compared to more standard Bayesian inference procedures, e.g., sampling approaches (Daunizeau et al., 2014).

In this note, we review the main variants of VL approaches, that follow from considering nonlinear models of continuous and/or categorical data. Note that most of the material we present here has already been described in previous (though separate) publications. What we hope to achieve with this report is to provide an exhaustive synthesis of the VL approach that would enable a plug-and-play implementation, without sacrificing too much mathematical rigour. *En passant*, we also derive a few novel theoretical results that complete the portfolio of existing analyses of variational Bayesian approaches, including investigations of their asymptotic convergence (Wang and Titterington, 2004; Westling and McCormick, 2015). For example, we show how the frequentist limit of VL estimators relate to classical maximum-likelihood estimators. We also suggest practical ways of extending existing VL approaches to hierarchical generative models that include (e.g., precision) hyperparameters (Goel and Degroot, 1981).

## 1. The variational Laplace approach

Let a generative model $m$ of data $y$ be defined in terms of both the likelihood function $p(y|\theta,m)$ and the prior density $p(\theta|m)$ on model parameters $\theta$. This immediately induces the following Free-Energy lower bound $F(q)$ on the log-model evidence $\log p(y|m)$:

$$\begin{aligned} F(q) &= \log p(y|m) - D_{KL}(q(\theta); p(\theta|y,m)) \\ &= \langle \log p(y|\theta,m) \rangle_{q(\theta)} - D_{KL}(q(\theta); p(\theta|m)) \\ &= \langle \log p(y|\theta,m) + \log p(\theta|m) \rangle_{q(\theta)} + S(q(\theta)) \end{aligned} \quad (1)$$

where $q(\theta)$ is an arbitrary probability density function over model parameters. Here, $S(\bullet)$ and $D_{KL}(\bullet;\bullet)$ denote the Shannon entropy and the Kullback-Leibler divergence, respectively. Note that Equation 1 summarizes three different perspectives on free energy. Given that Kullback-Leibler divergences are always positive, the first line in Equation 1 shows why the free energy lower bounds the log model evidence for any arbitrary probability density function $q(\theta)$. The second line captures the idea that free energy is a trade-off between fit accuracy (cf. expected log-likelihood) and model complexity (cf. divergence between the prior and the approximate posterior $q(\theta)$). The last line relates free energy to the objective function in max-entropic inference (Jaynes, 1957a, 1957b). In what follows, we will derive analytical expressions for the free energy bound from the latter expression, which will prove easier to manipulate.

Maximizing the free energy $F(q)$ with respect to $q(\theta)$ (under some simplifying constraints) is the basis for Variational Bayes (VB) inference approaches (Attias, 1999; Beal, 2003), and makes $q(\theta) \approx p(\theta|y,m)$ an approximate posterior density:

$$\frac{\delta F}{\delta q} = 0 \Rightarrow q(\theta) \propto \exp(I(\theta))$$
$$I(\theta) = \log p(y|\theta, m) + \log p(\theta|m)$$
(2)

where $I(\theta)$ is termed the *variational energy*. Note that Equation 2 is trivial, because no simplifying constraint was imposed on $q(\theta)$. Nevertheless, its usefulness is twofold: (i) it shows how Bayes' rule can be turned into an optimization problem, and (ii) it shows that VB is theoretically consistent with exact Bayesian inference.

Equation 2 can be generalized to less trivial cases (e.g. hierarchical generative models) that necessitate e.g., *mean-field* approximations to the posterior density $q(\theta)$ (Wainwright and Jordan, 2008). We will see examples of this below. Alternatively, the so-called *Variational Laplace* or VL approach (Daunizeau et al., 2009; Friston et al., 2007) uses a fixed-form Gaussian approximation to the posterior density, i.e.: $q(\theta) \triangleq N(\mu, \Sigma)$ is assumed to be Gaussian with first two moments $\mu$ and $\Sigma$ given by:

$$\begin{cases} \mu = \arg\max_{\theta} I(\theta) \\ \Sigma = -\left[ \frac{\partial^2 I}{\partial \theta^2} \bigg|_{\mu} \right]^{-1} \end{cases}$$
(3)

which basically relies upon a second-order Taylor expansion of the variational energy. Without loss of generality, this then yields the following expression for the free energy:

$$\begin{aligned} F(q) &= F(\mu, \Sigma) \\ &= \langle I(\theta) \rangle_q + \frac{1}{2} \log |\Sigma| + \frac{n_\theta}{2} \log 2\pi e \\ &\approx I(\mu) + \frac{1}{2} tr \underbrace{\left[ \frac{\partial^2 I}{\partial \theta^2} \bigg|_\mu \Sigma \right]}_{-n_\theta} + \frac{1}{2} \log |\Sigma| + \frac{n_\theta}{2} \log 2\pi e \\ &= I(\mu) + \frac{1}{2} \log |\Sigma| + \frac{n_\theta}{2} \log 2\pi \end{aligned} \quad (4)$$

Note that the passage from the third to the fourth line in Equation 4 follows from replacing $\Sigma$ with its expression in Equation 3.

One can check that the VL approach is consistent, i.e. the first- and second-order moments of the approximate posterior $q(\theta)$ maximize the free energy given in (the second line of) Equation 4:

$$(\mu^*, \Sigma^*) \triangleq \arg\max_{\mu, \Sigma} F(\mu, \Sigma) \Rightarrow \begin{cases} \frac{\partial F}{\partial \Sigma} \bigg|_{\mu^*, \Sigma^*} = 0 = \frac{1}{2} \frac{\partial^2 I}{\partial \theta^2} \bigg|_{\mu^*} + \frac{1}{2} \Sigma^{*-1} \Rightarrow \Sigma^* = -\left[ \frac{\partial^2 I}{\partial \theta^2} \bigg|_{\mu^*} \right]^{-1} \\ \frac{\partial F}{\partial \mu} \bigg|_{\mu^*, \Sigma^*} = 0 = \frac{\partial I}{\partial \theta} \bigg|_{\mu^*} \Rightarrow \mu^* = \arg\max_{\theta} I(\theta) \end{cases} \quad (5)$$

From Equation 5, one can see that the final expression for the Free Energy given in Equation 4 is only justified at convergence of the approach (i.e. when $\mu$ has been optimized).

Replacing the variational energy in Equation 4 with its definition in Equation 2 now yields:

$$F(q) \approx \underbrace{\log p(y|\mu, m)}_{\text{fit accuracy term}} + \underbrace{\log p(\mu|m) + \frac{1}{2} \log |\Sigma| + \frac{n_\theta}{2} \log 2\pi}_{\text{model complexity penalty term}} \quad (6)$$

where, by construction, the model complexity penalty term is what remains from removing the log likelihood (fit accuracy) term from the free energy. One can already see that this penalty term is not just a simple function of the number of model parameters: it depends upon the posterior density moments $\mu$ and $\Sigma$. In particular, this recapitulates the established idea that the "complexity" of a model is not an intrinsic property: it actually depends upon the fitted data (Spiegelhalter et al., 2002). We will further expand the expression of the model complexity penalty term in the next section.

## 2. (Conditionally) Gaussian observations

In this section, we will focus on the following class of nonlinear generative models $m$:

$$\left. \begin{array}{l} y = g(\theta) + \varepsilon_y \\ \varepsilon_y \sim N(0, Q) \end{array} \right\} \Rightarrow p(y|\theta, m) = N(g(\theta), Q) \qquad (7)$$

where the data $y$ are continuous variables (of dimension $n_y$), $\theta$ are unknown model parameters (of dimension $n_\theta$) observed through the mapping $g$ and $\varepsilon_y$ are Gaussian residuals with covariance structure $Q$. Equation 6 specifies the likelihood function $p(y|\theta, m)$ of the generative model $m$. In brief, conditional on model parameters $\theta$, observations $y$ behave as Gaussian variables with mean $g(\theta)$ and covariance $Q$.

### 2.1 The VL approach for continuous observation mappings

Without loss of generality, we assume that the generative model $m$ is equipped with Gaussian priors $p(\theta|m) \triangleq N(\mu_0, \Sigma_0)$ on model parameters. The variational energy $I(\theta)$ of Equation 2 then reduces to:

$$I(\theta) = -\frac{1}{2}\left(n_y \log 2\pi + \log|Q| + \varepsilon_y^T Q^{-1} \varepsilon_y + n_\theta \log 2\pi + \log|\Sigma_0| + \varepsilon_\theta^T \Sigma_0^{-1} \varepsilon_\theta\right) \quad (8)$$

where $\varepsilon_\vartheta = \mu_0 - \theta$ is the prediction error at the second level of the hierarchy. If the observation mapping $g$ is continuous, the covariance matrix of the approximate posterior density $q(\theta)$ is given by:

$$\Sigma = \left[\left.\frac{\partial g}{\partial \theta}\right|_\mu^T Q^{-1} \left.\frac{\partial g}{\partial \theta}\right|_\mu + \Sigma_0^{-1}\right]^{-1} \quad (9)$$

where derivatives of $g$ of order higher than one have been neglected, to ensure positive-definiteness of the posterior covariance matrix $\Sigma$.

Inserting Equation 8 into Equation 4 yields the following expression of the free energy $F$:

$$F = \underbrace{-\frac{1}{2}\left(n_y \log 2\pi + \log|Q| + \hat{\varepsilon}_y^T Q^{-1} \hat{\varepsilon}_y\right)}_{\text{fit accuracy term}} \underbrace{-\frac{1}{2}\left(\hat{\varepsilon}_\theta^T \Sigma_0^{-1} \hat{\varepsilon}_\theta + \log|\Sigma_0| - \log|\Sigma|\right)}_{\text{model complexity penalty term}}$$

$$\hat{\varepsilon}_y = y - g(\mu) \quad (10)$$
$$\hat{\varepsilon}_\theta = \mu_0 - \mu$$

Note that the model complexity penalty term has now a closed-form expression, which can be decomposed into two subcomponents. First, $\frac{1}{2}\hat{\varepsilon}_\theta^T \Sigma_0^{-1} \hat{\varepsilon}_\theta$ penalizes estimates that are *a priori* unlikely. This term vanishes at the frequentist limit, i.e. when $\Sigma_0 \to \infty$. Second, $\frac{1}{2}\log|\Sigma_0| - \frac{1}{2}\log|\Sigma|$ is equal to the (prior to posterior) reduction of uncertainty on model parameters. Taken together, (prior to posterior) *changes in the information about model parameters will be seen as a signature of model complexity and will be penalized accordingly.* Note that this intuition generalizes to any form of generative model, including those reviewed below (cf. multinomial likelihood functions, etc...).

Equations 3, 9 and 10 summarize the standard VL approach for non-hierarchical generative models of continuous data, which is used to finesse the inversion of nonlinear models with (conditionally) Gaussian likelihoods (Daunizeau et al., 2009; Friston et al., 2007).

### 2.2 Mean-field separation with precision hyperparameters

Let us now consider a hierarchical extension to the above generative model, which accounts for uncertainty in the second-order moments of both the likelihood and the prior. This is typically done by augmenting the generative model with (unknown) precision hyperparameters $\lambda$ that determine the prior covariance matrices of both the noise and parameters, as follows (Dempster et al., 1981; Rao, 1972):

$$\begin{aligned}
Q^{-1} &\equiv Q^{-1}(\lambda_y) \\
&\approx \sum_i \lambda_y^{(i)} \underbrace{\frac{\partial Q^{-1}}{\partial \lambda_y^{(i)}}}_{\Phi_y^{(i)}} \\
\Sigma_0^{-1} &\equiv \Sigma_0^{-1}(\lambda_\theta) \\
&\approx \sum_i \lambda_\theta^{(i)} \underbrace{\frac{\partial \Sigma_0^{-1}}{\partial \lambda_\theta^{(i)}}}_{\Phi_\theta^{(i)}}
\end{aligned} \qquad (11)$$

where $\Phi_y^{(i)}$ and $\Phi_\theta^{(i)}$ form a fixed covariance matrix basis set. Equation 11 can be used to capture any covariance structure. At the limit, one could use one hyperpameter per degree of freedom in the covariance matrices. More practically, one would typically use a basis set composed of positive-definite matrices, and then place either Gamma or log-Normal priors $p(\lambda|m)$ on $\lambda$ to enforce positivity constraints (see e.g., Friston et al., 2008).

An effective VB approach to the ensuing inference problem relies upon a mean-field separation of posterior densities on $\theta$ and $\lambda$ (Giordano et al., 2015). More precisely, one is looking for the marginal densities $q(\theta)$ and $q(\lambda)$ that maximize the free energy $\tilde{F}$ under the separability constraint $q(\theta,\lambda) = q(\theta)q(\lambda)$. It turns out that one can still use the above VL approach for $\theta$, by replacing the variational energy $I(\theta)$ given in Equation 8 by the variational energy $\tilde{I}_\theta(\theta)$ that results from zeroing the gradient of the free energy $\tilde{F}$:

$$\begin{aligned}
\frac{\delta \tilde{F}}{\delta q(\theta)} &= 0 \Rightarrow q(\theta) \propto \exp \tilde{I}_\theta(\theta) \\
\tilde{I}_\theta(\theta) &\equiv \left\langle \tilde{L}(\theta,\lambda) \right\rangle_{q(\lambda)} \\
&= -\frac{1}{2} n_y \log 2\pi - \frac{1}{2} n_y \left\langle \varepsilon_y^T Q^{-1}(\lambda_y) \varepsilon_y - \log \left| Q^{-1}(\lambda_y) \right| \right\rangle_{q(\lambda)} \\
&\quad -\frac{1}{2} n_\theta \log 2\pi - \frac{1}{2} \left\langle \varepsilon_\theta^T \Sigma_0^{-1}(\lambda_\theta) \varepsilon_\theta - \log \left| \Sigma_0^{-1}(\lambda_\theta) \right| \right\rangle_{q(\lambda)} \\
&\quad + \left\langle \log p(\lambda|m) \right\rangle_{q(\lambda)}
\end{aligned} \quad (12)$$

where $\tilde{L}(\theta,\lambda) = \log p(y|\theta,\lambda m) + \log p(\theta|m) + \log p(\lambda|m)$ is the log-joint density of the generative model, having included the precision hyperparameters $\lambda$.

Note that when only one covariance component per variable set is used (i.e. when $Q^{-1} = \lambda_y \Phi_y$ and $\Sigma_0^{-1} = \lambda_\theta \Phi_\theta$), the variational energy $\tilde{I}_\theta(\theta)$ in Equation 12 simplifies to:

$$\begin{aligned}
\tilde{I}_\theta(\theta) &= -\frac{1}{2}\left( n_y \log 2\pi + \left\langle \lambda_y \right\rangle \varepsilon_y^T \Phi_y \varepsilon_y - \ln \left| \Phi_y \right| + \log 2\pi + \left\langle \lambda_\theta \right\rangle \varepsilon_\theta^T \Phi_\theta \varepsilon_\theta - \log \left| \Phi_\theta \right| \right) \\
&\quad + \frac{n_y}{2} \left\langle \log \lambda_y \right\rangle + \frac{n_\theta}{2} \left\langle \log \lambda_\theta \right\rangle + \left\langle \log p(\lambda|m) \right\rangle \\
&= I(\theta) + \frac{n_y}{2} \left( \left\langle \log \lambda_y \right\rangle - \log \left\langle \lambda_y \right\rangle \right) + \frac{n_\theta}{2} \left( \left\langle \log \lambda_\theta \right\rangle - \log \left\langle \lambda_\theta \right\rangle \right) + \left\langle \log p(\lambda|m) \right\rangle
\end{aligned} \quad (13)$$

where $I(\theta)$ improperly refers to the variational energy of $\theta$ as given in Equation 8, having replaced the precision matrices $Q^{-1}$ and $\Sigma_0^{-1}$ with their VB estimate $\hat{Q}^{-1} = \langle \lambda_y \rangle \Phi_y$ and $\hat{\Sigma}_0^{-1} = \langle \lambda_\theta \rangle \Phi_\theta$, respectively. It follows that the free energy $\tilde{F}$ that ensues from considering uncertain precision hyperparameters $\lambda$ can be written as the free energy $F$ of the model with "fixed" covariances $\hat{Q}^{-1}$ and $\hat{\Sigma}_0^{-1}$ plus a correction term $\Delta F$:

$$\begin{aligned}\tilde{F} &= \langle L(\theta,\lambda) \rangle_{q(\theta)q(\lambda)} - \langle \log q(\theta) \rangle_{q(\theta)} - \langle \log q(\lambda) \rangle_{q(\lambda)} \\ &\approx \underbrace{I(\mu) + \frac{1}{2}\log|\Sigma| + \frac{n_\theta}{2}\log 2\pi}_{F} + \Delta F\end{aligned} \quad (14)$$

where $I(\mu)$ refers to $I(\theta)$ evaluated at the mode $\mu = \langle \theta \rangle_{q(\theta)}$ of the approximate posterior $q(\theta)$, and the correction term $\Delta F$ depends upon the form of the prior and posterior densities on precision hyperparameters $\lambda$ (see below).

Without loss of generality, let us consider the following Gamma priors on hyperparameters:

$$\begin{cases} p(\lambda_y | m) = Ga(a_y^0, b_y^0) \\ \qquad = \frac{1}{\Gamma(a_y^0)} \lambda_y^{a_y^0 - 1} \exp(-b_y^0 \lambda_y + a_y^0 \ln b_y^0) \\ p(\lambda_\theta | m) = Ga(a_\theta^0, b_\theta^0) \\ \qquad = \frac{1}{\Gamma(a_\theta^0)} \lambda_\theta^{a_\theta^0 - 1} \exp(-b_\theta^0 \lambda_\theta + a_\theta^0 \ln b_\theta^0) \end{cases} \quad (15)$$

where $(a_y^0, b_y^0)$ (resp. $(a_\theta^0, b_\theta^0)$) are the scale and shape parameters of the prior Gamma density on the noise (resp. parameters) precision $\lambda_y$ (resp. $\lambda_\theta$). Note that so-called reference *noninformative* priors $p(\lambda) \propto \lambda^{-1}$ are a special case of Equation 15, with $a^0 = b^0 = 0$ (Jeffreys, 1946).

The variational energy $\tilde{I}_\lambda(\lambda)$ of hyperparameters then obtains from zeroing the functional gradient of the free energy w.r.t. to $q(\lambda)$:

$$\begin{aligned}
\frac{\delta \tilde{F}}{\delta q(\lambda)} &= 0 \Rightarrow q(\lambda) \propto \exp \tilde{I}_\lambda(\lambda) \\
\tilde{I}_\lambda(\lambda) &\equiv \langle \tilde{L}(\theta,\lambda) \rangle_{q(\theta)} \\
&= -\frac{1}{2}\lambda_y \left( \hat{\varepsilon}_y^T \Phi_y \hat{\varepsilon}_y + tr\left[ \left.\frac{\partial g}{\partial \theta}\right|_\mu^T \Phi_y \left.\frac{\partial g}{\partial \theta}\right|_\mu \Sigma \right] \right) + \frac{n_y}{2}\log \lambda_y \\
&\quad -\frac{1}{2}\lambda_\theta \left( \hat{\varepsilon}_\theta^T \Phi_\theta \hat{\varepsilon}_\theta + tr[\Phi_\theta \Sigma] \right) + \frac{n_\theta}{2}\log \lambda_\theta + \log p(\lambda|m) + cst
\end{aligned} \quad (16)$$

One can see from Equation 16 that the variational energy $\tilde{I}_\lambda(\lambda)$ contains no interaction term between $\lambda_y$ and $\lambda_\theta$, which implies that their approximate marginal densities are separable. Also, $\tilde{I}_\lambda(\lambda)$ has the form of a sum of two (log-) Gamma densities. This is important, because it greatly simplifies the VB update rule for $\lambda$, which reduces to:

$$\begin{cases} q(\lambda_y) = Ga(a_y, b_y) \\ q(\lambda_\theta) = Ga(a_\theta, b_\theta) \end{cases} \Rightarrow \begin{cases} a_y = a_y^0 + \dfrac{n_y}{2} \\ b_y = b_y^0 + \dfrac{1}{2}\left( \hat{\varepsilon}_y^T \Phi_y \hat{\varepsilon}_y + tr\left[ \left.\dfrac{\partial g}{\partial \theta}\right|_\mu^T \Phi_y \left.\dfrac{\partial g}{\partial \theta}\right|_\mu \Sigma \right] \right) \\ a_\theta = a_\theta^0 + \dfrac{n_y}{2} \\ b_\theta = b_\theta^0 + \dfrac{1}{2}\left( \hat{\varepsilon}_\theta^T \Phi_\theta \hat{\varepsilon}_\theta + tr[\Phi_\theta \Sigma] \right) \end{cases} \quad (17)$$

The ensuing free energy correction term $\Delta F$ is then given by:

$$\begin{aligned}
\Delta F &= \frac{n_y}{2}\left( \langle \log \lambda_y \rangle - \log\langle \lambda_y \rangle \right) + \frac{n_\theta}{2}\left( \langle \log \lambda_\theta \rangle - \log\langle \lambda_\theta \rangle \right) - D_{KL}(q(\lambda); p(\lambda|m)) \\
&= \sum_{x \in (y,\theta)} a_x^0 \log \frac{b_x^0}{b_x} - \frac{n_x}{2}\log a_x - \log \frac{\Gamma(a_x^0)}{\Gamma(a_x)} + a_x\left(1 - \frac{b_x^0}{b_x}\right)
\end{aligned} \quad (18)$$

where the sum is over variable sets.

Iterating over the VL scheme (cf. Equations 5-9) and the VB update rule for $\lambda$ (Equation 17) eventually provides approximate posterior densities over both model parameters ($\theta$) and hyperparameters ($\lambda$), as well as a lower bound to the model evidence $\tilde{F}$. Note that adjusting prior hyperparameters $\lambda_\theta$ is the hallmark of so-called "empirical Bayes" approaches (Robbins, 1964), which possess desirable statistical properties (Carlin and Louis, 1997; Efron and Morris, 1973).

### 2.3 The frequentist limit for linear models

This section revisits the above VB algorithm in the context of a linear observation mapping (cf. general linear models or GLMs) at the frequentist limit, i.e. when priors on $\theta$ become noninformative (Datta and Ghosh, 1995).

First of all, note that for a linear observation mapping (i.e.: $g(\theta) = X\theta$, where $X$ is any arbitrary $n_y \times n_\theta$ matrix), it is easy to prove that the VB update rules (under the above mean-field partition) on both the parameters and hyperparameters write:

$$q(\theta): \begin{cases} \mu = \mu_0 + \langle \lambda_y \rangle \Sigma X^T \Phi_y (y - X\mu_0) \\ \Sigma = \left( \langle \lambda_y \rangle X^T \Phi_y X + \Sigma_0^{-1} \right)^{-1} \end{cases}$$

$$q(\lambda_y): \begin{cases} a_y = a_y^0 + \dfrac{n_y}{2} \\ b_y = \dfrac{b_y^0 + \dfrac{1}{2}(y - X\mu)^T \Phi_y (y - X\mu)}{1 - \dfrac{n_\theta}{2a_y^0 + n_y}\left(1 - tr\left[\Sigma_0^{-1}\Sigma\right]\right)} \end{cases} \quad (19)$$

where $\Phi_y$ is the fixed precision component and $\langle \lambda_y \rangle = a_y / b_y$ by virtue of the Gamma density parameterization. Note that the last line of Equation 19 derives from noting that, if $X$ is full-rank, then $tr\left[X^T \Phi_y X \Sigma\right] = n_\theta \langle \lambda_y \rangle^{-1} \left(1 - tr\left[\Sigma_0^{-1} \Sigma\right]\right)$.

Usually, VB update rules are iterated until the free energy reaches a plateau. However, VB estimates at the frequentist limit does not require any iteration, and can be evaluated analytically from Equation 19:

$$\mu \xrightarrow{\Sigma_0 \to \infty} \left(X^T \Phi_y X\right)^{-1} X^T \Phi_y y$$
$$\Sigma \xrightarrow{\Sigma_0 \to \infty} \langle \lambda_y \rangle^{-1} \left(X^T \Phi_y X\right)^{-1}$$
$$a_y \xrightarrow{a_y^0 \to 0} \frac{n_y}{2} \quad (20)$$
$$b_y \xrightarrow{b_y^0 \to 0, \Sigma_0 \to \infty} \frac{1}{2} \frac{n_y}{n_y - n_\theta} (y - X\mu)^T \Phi_y (y - X\mu)$$

In particular, Equation 20 yields the following VB estimate $\langle \lambda_y \rangle^{-1}$ of the residual noise variance:

$$\langle \lambda_y \rangle^{-1} \xrightarrow{\Sigma_0 \to \infty} \frac{(y - X\mu)^T \Phi_y (y - X\mu)}{n_y - n_\theta} \quad (21)$$

Interestingly, Equations 20 and 21 yield VB estimators that are equal to the maximum likelihood estimates. The loss of degrees of freedom in the noise variance estimate (cf. numerator of Equation 21) follows from having properly accounted for the uncertainty about $\theta$ when updating the moments of the approximate posterior on the noise variance $q(\lambda_y)$. This exemplifies the consistency of VL estimators at the frequentist limit.

Finally, note that, at the frequentist limit, the model evidence can be evaluated in closed form. To begin with, it can be shown that:

Notes on the Laplace approximation. J. Daunizeau, 2011.$$(y-X\theta)^T \Phi_y (y-X\theta) = (\mu-\theta)^T (X^T \Phi_y X)(\mu-\theta) + y^T P y \qquad (22)$$

where $P$ is a projector defined as: $P = \Phi_y - \Phi_y X (X^T \Phi_y X)^{-1} X^T \Phi_y$. This will prove useful when deriving a closed-form expression for the marginal likelihood:

$$\begin{aligned}
p(y|m) &= \iint p(y|\theta,\lambda,m) \lambda^{-1} d\theta d\lambda \\
&= \int (2\pi)^{-\frac{n_y}{2}} |\Phi_y|^{\frac{1}{2}} \lambda^{\frac{n_y}{2}-1} d\lambda \int \exp\left[-\frac{\lambda}{2}(y-X\theta)^T \Phi_y (y-X\theta)\right] d\theta \\
&= (2\pi)^{-\frac{n_y}{2}} |\Phi_y|^{\frac{1}{2}} \int \lambda^{\frac{n_y}{2}-1} \exp\left[-\frac{\lambda}{2} y^T P y\right] d\lambda \underbrace{\int \exp\left[-\frac{\lambda}{2}(\mu-\theta)^T (X^T \Phi_y X)(\mu-\theta)\right] d\theta}_{(2\pi)^{\frac{n_\theta}{2}} |\lambda X^T \Phi_y X|^{-\frac{1}{2}}} \\
&= (2\pi)^{\frac{n_\theta - n_y}{2}} |\Phi_y|^{\frac{1}{2}} |X^T \Phi_y X|^{-\frac{1}{2}} \underbrace{\int \lambda^{\frac{n_y - n_\theta}{2}-1} \exp\left[-\frac{\lambda}{2} y^T P y\right] d\lambda}_{\Gamma\left(\frac{n_y - n_\theta}{2}\right) \big/ \left(\frac{y^T P y}{2}\right)^{\frac{n_y - n_\theta}{2}}} \\
&= (2\pi)^{\frac{n_\theta - n_y}{2}} |\Phi_y|^{\frac{1}{2}} |X^T \Phi_y X|^{-\frac{1}{2}} \Gamma\left(\frac{n_y - n_\theta}{2}\right) \left(\frac{y^T P y}{2}\right)^{\frac{n_\theta - n_y}{2}}
\end{aligned}$$

(23)

where the passage from the third to the fourth line follows from the functional form of the multivariate normal density and the passage from the fourth to the fifth line follows from the functional form of the Gamma density.

Recall that, formally speaking, the frequentist limit to the Laplace free energy is unbounded (i.e.: $\tilde{F} \xrightarrow{\Sigma_0 \to \infty} -\infty$), because it induces improper prior distributions on model parameters (cf. prior entropy terms $-\frac{1}{2} \ln |\Sigma_0|$ in Equation 10 and $-\log \Gamma(a_{y/\theta}^0)$ from Equation 18). However, model comparison proceeds from free energy differences. Those should be immune to the (badly-behaved) prior entropy terms, which, at the frequentist limit, are the same for all models. We will thus remove these terms when replacing Equation 20 into Equations 14-18, to yield a pseudo frequentist limit $F_\infty \neq \lim_{\Sigma_0 \to \infty} \tilde{F}$ to the Laplace free energy:

$$\begin{aligned}F_\infty &= -\frac{n_y}{2}\log 2\pi + \frac{1}{2}\log|\Phi_y| - \frac{1}{2}\log|X^T \Phi_y X| - \frac{n_y - n_\theta}{2}\log\frac{y^T P y}{2} \\ &\quad + \frac{n_y - n_\theta}{2}\log\left(\frac{n_y - n_\theta}{2}\right) - \frac{n_y}{2}\log\frac{n_y}{2} + \log\Gamma\left(\frac{n_y}{2}\right) + \frac{n_\theta}{2} \\ &= \log p(y|m) + \frac{n_\theta}{2}(1 - \log 2\pi) + \frac{n_y - n_\theta}{2}\log\left(\frac{n_y - n_\theta}{2}\right) - \frac{n_y}{2}\log\frac{n_y}{2} - \log\Gamma\left(\frac{n_y - n_\theta}{2}\right) + \log\Gamma\left(\frac{n_y}{2}\right) \\ &\approx \log p(y|m) - \frac{1}{2}n_\theta \log 2\pi + \frac{1}{2}\log\frac{n_y}{n_y - n_\theta}\end{aligned}$$

(24)

where the last line derives from a variant of Stirling's approximation (Marsaglia and Marsaglia, 1990), i.e.: $\log\Gamma(x) - x\log x \approx -\frac{1}{2}\log x - x + \frac{1}{2}\log 2\pi$, which will be accurate at high sample sizes (i.e. when $n_y \gg n_\theta$). Numerical simulations confirm this approximation in showing that for most sample sizes $n_y$, the difference $F_\infty - \log p(y|m)$ behaves as an affine function of $n_\theta$, with a near unit slope (recall that $\frac{1}{2}\log 2\pi \approx 0.92$) and an offset $O(n_y)$ that depends upon $n_y$, i.e.: $F_\infty \approx \log p(y|m) - 0.92 n_\theta + O(n_y)$. This is exemplified on Figure 1 below. Note that the difference between the pseudo frequentist limit to the Laplace free energy and the log marginal likelihood depends neither upon high-level statistical features of the generative model (e.g. the design matrix $X$ or the noise covariance structure $\Phi_y$), nor upon the observed data sample $y$. However, it depends upon the model dimensions, in particular: the number of model parameters $n_\theta$. This means that the effective complexity penalty term of $F_\infty$ is partly confounded by the improper priors induced by the frequentist limit. This also implies that model comparisons based upon these two metrics will not yield identical results.

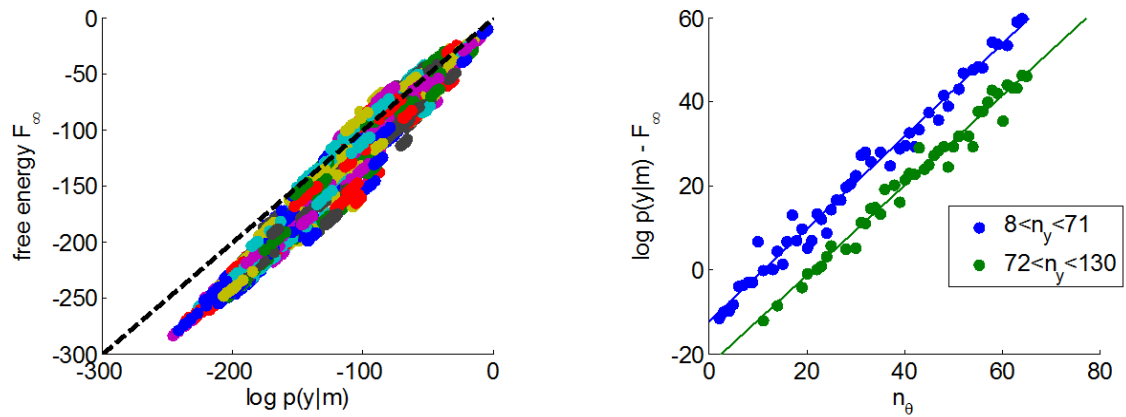

**Figure 1: comparison of the pseudo frequentist limit to the Laplace free energy and the log marginal likelihood**. Dummy data were simulated under random general linear models with varying model dimensions ($n_\theta$ and $n_y$). For each simulated data, both $F_\infty$ and $\log p(y|m)$ were evaluated. **Left**: $F_\infty$ (y-axis) is plotted as a function of $\log p(y|m)$ (y-axis) for each simulated dataset. Each colour indicates a specific pair of $n_\theta$ and $n_y$. The black dotted line indicates the main diagonal, i.e. the ideal identity mapping $F_\infty = \log p(y|m)$. One can see that the variational Laplace approach arguably yields a reasonable approximation to the log model evidence. However, the cloud of $F_\infty$ points does not really scatter around the main diagonal: the difference between the two gets larger as $\log p(y|m)$ decreases. In addition, one can see that the pseudo frequentist limit $F_\infty$ is not a strict lower bound on $\log p(y|m)$, i.e. the difference $\log p(y|m) - F_\infty$ can become negative at times. This is due to the heuristic removal of prior entropy terms. **Right**: the mean difference $\log p(y|m) - F_\infty$ (y-axis) is plotted as a function of $n_\theta$ (y-axis) for both small ($8 < n_y < 71$, blue dots) and big ($72 < n_y < 130$, green dots) sample sizes, according to a median-split among simulated datasets. The plain lines show the best fitting linear trends, which have near unit slopes.

# 3. Variational Laplace with multinomial observations

The VL treatment above copes with continuous data, for which there is a natural distance metric. Now if the data is categorical, there is no such natural metric, and one has to resort to probability distributions dealing with discrete events. The most generic case is then to consider multinomial likelihood functions, under which the Laplace approach can still be derived. This is the focus of the current section.

**3.1 The Bernouilli case**

First, let us consider simple i.i.d. Bernouilli observations:

$$p(y|\theta,m) = \prod_{i=1}^{n_y} p(y_i|\theta,m)$$
$$p(y_i|\theta,m) = g_i(\theta)^{y_i}(1-g_i(\theta))^{1-y_i}$$
(25)

where $y_i$ is a binary data point (there are $n_y$ of these), whose first-order moment $g_i(\theta)$ is determined by model parameters, i.e.: $p(y_i=1|\theta,m) = \langle y_i|\theta,m\rangle = g_i(\theta)$. By analogy with the conditionally Gaussian model, we will refer to the vector-value function $g = \begin{bmatrix} g_1 & \cdots & g_{n_y} \end{bmatrix}^T$ as the observation mapping.

As above, given priors $p(\theta|m)$ on model parameters, any arbitrary probability density function $q(\theta)$ induces a free-energy bound $F$ on the log-model evidence $\log p(y|m)$:

$$F = \langle I(\theta) \rangle_{q(\theta)} + S(q(\theta))$$
$$\approx I(\mu) + \frac{1}{2}\ln|\Sigma| + \frac{n_\theta}{2}\ln 2\pi \qquad (26)$$
$$I(\theta) = \log p(\theta|m) + \sum_{i=1}^{n_y} y_i \log g_i(\theta) + (1-y_i)\log(1-g_i(\theta))$$

where we have used the Laplace approximation to go from the first to the second line, and the expression for the variational energy $I(\theta)$ derives from the definition of the Bernouilli likelihood in equation 25.

Here again, the VL approach relies upon a second-order Taylor expansion of the variational energy of model parameters, yielding a Gaussian fixed-form approximate posterior density $q(\theta) \triangleq N(\mu, \Sigma)$. Under Normal priors ($p(\theta|m) \triangleq N(\mu_0, \Sigma_0)$), the gradient and Hessian of the variational energy given in Equation 26 then write:

$$\frac{\partial I}{\partial \theta} = \Sigma_0^{-1}(\mu_0 - \theta) + \sum_{i=1}^{n_y} \frac{y_i - g_i(\theta)}{g_i(\theta)(1-g_i(\theta))} \frac{\partial g_i}{\partial \theta}$$
$$\frac{\partial^2 I}{\partial \theta^2} \approx -\Sigma_0^{-1} - \sum_{i=1}^{n_y}\left(\frac{y_i}{g_i(\theta)^2} + \frac{1-y_i}{(1-g_i(\theta))^2}\right)\frac{\partial g_i}{\partial \theta}\frac{\partial g_i}{\partial \theta}^T \qquad (27)$$

where the second-order derivatives of the observation mapping have been neglected. This is actually necessary in the general case, because it can potentially induce non positive-definite covariance matrices (but see the sigmoid case below). Inserting Equation 27 into Equation 3 yields an analytical form for the second-order moment of the approximate posterior density $q(\theta)$:

$$\Sigma \approx \left[\Sigma_0^{-1} + \frac{\partial g}{\partial \theta} Diag\left(\frac{y}{g(\theta)^2} + \frac{1-y}{(1-g(\theta))^2}\right)\frac{\partial g}{\partial \theta}^T\right]^{-1} \qquad (28)$$

where the divide and square operators within the brackets of the right-hand-side of Equation 28 improperly denote the element-by-element operations on the appropriate vectors.

There is a very frequent special case of observation mappings that greatly simplifies the VL treatment of Bernouilli data, namely a linear mixture of $\theta$ passed through a sigmoid mapping:

$$g_i : \theta \to g_i(\theta) = \frac{1}{1+\exp(-A_i^T \theta + b_i)} \Rightarrow \begin{cases} \frac{\partial I}{\partial \theta} = \Sigma_0^{-1}(\mu_0 - \theta) + \sum_{i=1}^{n_y}(y_i - g_i(\theta))A_i \\ \frac{\partial^2 I}{\partial \theta^2} = -\Sigma_0^{-1} - \sum_{i=1}^{n_y} g_i(\theta)(1 - g_i(\theta)) A_i A_i^T \end{cases} \quad (29)$$

$$\Rightarrow \Sigma = \left[ \Sigma_0^{-1} + A\, Diag\big(g(\theta)(1-g(\theta))\big) A^T \right]^{-1}$$

where $A_i$ is an arbitrary vector of same size than $\theta$ and $b_i$ is a scalar. In this case, there no first-order approximation for the Hessian of the variational energy is required. Note that the generative model induced by Equation 29 is formally equivalent to logistic regression, which justifies our apparently restricted interest...

### 3.2 The binomial extension

In short, the Binomial distribution measures the frequency of outcomes induced by $k$ repetitions of Bernoulli trials (with identical probability $g$). The probability of getting exactly $y$ successes in $k$ trials is given by the following (binomial) probability mass function:

$$p(y|k,g) = \frac{k!}{y!(k-y)!} g^y (1-g)^{k-y} \quad (30)$$

where $y$ is a scalar that can take values between 0 and $k$.

As above, we will consider that (i) some parameters $\theta$ determine the Bernouilli probability $g$ (i.e.: $g \equiv g(\theta)$), and that (ii) the binomial parameters can change from one observation to the next (i.e.: the data point $y_i$ follows a binomial distribution with parameters $k_i$ and $g_i(\theta)$). Then the data likelihood can be written as follows:

$$p(y|\theta,m) = \prod_{i=1}^{n_y} p(y_i|\theta,m)$$
$$p(y_i|\theta,m) = \frac{k_i!}{y_i!(k_i-y_i)!} g_i(\theta)^{y_i} (1-g_i(\theta))^{k_i-y_i} \tag{31}$$

Under Normal priors ($p(\theta|m) \triangleq N(\mu_0, \Sigma_0)$), the variational energy $I(\theta)$ and its gradient and Hessian are given by:

$$I(\theta) = \log p(\theta|m) + \sum_{i=1}^{n_y} y_i \log g_i(\theta) + (k_i - y_i)\log(1-g_i(\theta)) + \sum_{i=1}^{n_y} \log \frac{k_i!}{y_i!(k_i-y_i)!}$$
$$\frac{\partial I}{\partial \theta} = \Sigma_0^{-1}(\mu_0 - \theta) + \sum_{i=1}^{n_y} \frac{y_i - k_i g_i(\theta)}{g_i(\theta)(1-g_i(\theta))} \frac{\partial g_i}{\partial \theta} \tag{32}$$
$$\frac{\partial^2 I}{\partial \theta^2} \approx -\Sigma_0^{-1} - \sum_{i=1}^{n_y} \left( \frac{y_i}{g_i(\theta)^2} + \frac{k_i - y_i}{(1-g_i(\theta))^2} \right) \frac{\partial g_i}{\partial \theta} \frac{\partial g_i}{\partial \theta}^T$$

which can directly be used to derive the first- and second-order moment of the approximate (Laplace) posterior density (according to Equation 3) and the ensuing free energy (according to Equation 4).

Note that here again, setting $g$ to be a linear mixture of $\theta$ passed through a sigmoid mapping, yields a drastic simplification to Equation 32:

$$g_i : \theta \to g_i(\theta) = \frac{1}{1+\exp(-A_i^T \theta + b_i)} \Rightarrow \begin{cases} \frac{\partial I}{\partial \theta} = \Sigma_0^{-1}(\mu_0 - \theta) + \sum_{i=1}^{n_y}(y_i - k_i\, g_i(\theta)) A_i \\ \frac{\partial^2 I}{\partial \theta^2} = -\Sigma_0^{-1} - \sum_{i=1}^{n_y} k_i\, g_i(\theta)(1 - g_i(\theta)) A_i A_i^T \end{cases} \quad (33)$$

$$\Rightarrow \Sigma = \left[\Sigma_0^{-1} + A\,\text{Diag}\left(k \times g(\theta) \times (1 - g(\theta))\right) A^T\right]^{-1}$$

where $\times$ denotes element-by-element product, $A_i$ is an arbitrary vector of same size than $\theta$ and $b_i$ is a scalar.

### 3.3. Multinomial data

The multinomial distribution is a generalization of the binomial distribution, which allows for more than two (mutually exclusive) outcomes in the canonical Bernouilli trial. The probability mass function of the multinomial distribution writes:

$$p(y|k,g) = \begin{cases} \dfrac{k!}{\prod_{j=1}^m y_j!} \prod_{j=1}^m g_j^{y_j} & \text{if } k = \sum_{j=1}^m y_j \\ 0 & \text{otherwise} \end{cases} \quad (34)$$

where $m$ is the number of possible outcomes. Here, $y$ is a $m \times 1$ vector, whose elements can take values between 1 and $k$, $g$ is a $m \times 1$ vector, whose elements give the frequency of the corresponding outcomes and $k$ is the number of "trials" consisting of drawing a sample from the $m \times 1$ categorical (canonical) distribution determined by $g$.

As for binomial data, we will consider that (i) some parameters $\theta$ determine the sufficient statistics $g_j$ (i.e.: $g_j \equiv g_j(\theta)$), and that (ii) the multinomial parameters can change from one observation to the next (i.e.: the data point $y_i$ follows a binomial distribution with parameters $k_i$ and $g_i(\theta)$). The likelihood function can then be written as follows:

$$p(y|\theta,m) = \prod_{i=1}^{n_y} p(y_i|\theta,m)$$
$$p(y_i|\theta,m) = \frac{k_i!}{\prod_{j=1}^{m} y_{ij}!} \prod_{j=1}^{m} g_{ij}(\theta)^{y_{ij}} \qquad (35)$$

where we have assumed that the normalization conditions $k = \sum_{j=1}^{m} y_{ij}$ and $1 = \sum_{j=1}^{m} g_{ij}(\theta)$ are met for all $i$. Note that the second condition can always be enforced by construction of the observation function $g$, using, e.g., softmax mappings (Daunizeau, 2017).

Under Normal priors ($p(\theta|m) \triangleq N(\mu_0, \Sigma_0)$), the variational energy $I(\theta)$ and its gradient and Hessian are given by:

$$I(\theta) = \log p(\theta|m) + \sum_{i=1}^{n_y} \sum_{j=1}^{m} y_{ij} \log g_{ij}(\theta) + \sum_{i=1}^{n_y} \left( \log k_i! - \sum_{j=1}^{m} \log y_{ij}! \right)$$
$$\frac{\partial I}{\partial \theta} = \Sigma_0^{-1}(\mu_0 - \theta) + \sum_{i=1}^{n_y} \sum_{j=1}^{m} \frac{y_{ij}}{g_{ij}(\theta)} \frac{\partial g_{ij}}{\partial \theta} \qquad (36)$$
$$\frac{\partial^2 I}{\partial \theta^2} \approx -\Sigma_0^{-1} - \sum_{i=1}^{n_y} \sum_{j=1}^{m} \frac{y_{ij}}{g_{ij}(\theta)^2} \frac{\partial g_{ij}}{\partial \theta} \frac{\partial g_{ij}}{\partial \theta}^T$$

Equation 36 can be used to derive the first- and second-order moment of the approximate (Laplace) posterior density (according to Equation 3) and the free energy (according to Equation 4). It can be seen that Equation 36 reduces to Equation 32 (binomial case) when considering $m=2$ outcomes and re-inserting the normalization constraint of Equation 31.

Also, Equation 30 itself reduces to Equation 20 (Bernouilli case) when setting the number of "trials" to $k=1$ for all $i$.

### 3.4 Mean-field separation with precision hyperparameters.

As we have seen already, one can account for uncertainty in the second-order moment of the prior $p(\theta|m)$ by augmenting the generative model with (unknown) precision hyperparameters $\lambda_\theta$ that determine the prior covariance matrix of the parameters. Here again, one would typically use a mean-field separation assumption along with the above VL approach for $\theta$ by replacing the variational energy $I(\theta)$ in Equation 34 with $\tilde{I}_\theta(\theta)$:

$$\frac{\delta \tilde{F}}{\delta q(\theta)} = 0 \Rightarrow q(\theta) \propto \exp \tilde{I}_\theta(\theta)$$

$$\tilde{I}_\theta(\theta) \equiv \sum_{i=1}^{n_y}\sum_{j=1}^{m} y_{ij} \log g_{ij}(\theta) + \sum_{i=1}^{n_y}\left(\log k_i! - \sum_{j=1}^{m} \log y_{ij}!\right)$$
$$-\frac{1}{2}n_\theta \log 2\pi - \frac{1}{2}\left\langle \varepsilon_\theta^T \Sigma_0^{-1}(\lambda_\theta)\varepsilon_\theta - \log\left|\Sigma_0^{-1}(\lambda_\theta)\right|\right\rangle_{q(\lambda)}$$
$$+\left\langle \log p(\lambda|m)\right\rangle_{q(\lambda)}$$

(37)

and applying the same correction term $\Delta F$, which is given by Equation 18 when using a Gamma prior on the hyperparameter $\lambda \equiv \lambda_\theta$.

## 4. Conclusion

In this note, we have reviewed the VL approach to approximate Bayesian inference on nonlinear models of continuous and/or categorical data.

All the VL variants that we have described here are implemented in the VBA freeware (Daunizeau et al., 2014): http://mbb-team.github.io/VBA-toolbox/. This also includes extended VL approaches that deal with unknown precision hyperparameters.